\documentclass[
  ,final            
  ]
  {aipproc}

\usepackage{color}
\usepackage{graphicx}
\usepackage{caption}
\usepackage{subcaption}

\layoutstyle{6x9}

\newcommand{\CHAIN}[1]{\mathbf{#1}}
\renewcommand{\P}{{\CHAIN{P}}}
\newcommand{\Q}{{\CHAIN{Q}}}

\begin{document}

\title{Information-Based Physics, Influence, and Forces}



\author{James Lyons Walsh}{
  address={Department of Physics, University at Albany (SUNY)}
}

\author{Kevin H. Knuth}{
  address={Department of Physics, University at Albany (SUNY)}
  ,altaddress={Department of Informatics, University at Albany (SUNY)}}

\begin{abstract}

In recent works, Knuth and Bahreyni have demonstrated that the concepts of space and time are emergent in a coarse-grained model of direct particle-particle influence.  In addition, Knuth demonstrated that observer-made inferences regarding the free particle, which is defined as a particle that influences others, but is not itself influenced, result in a situation identical to the Feynman checkerboard model of the Dirac equation.  This suggests that the same theoretical framework that gives rise to an emergent spacetime is consistent with quantum mechanics. In this paper, we begin to explore the effect of influence on the emergent properties of a particle.  This initial study suggests that when a particle is influenced, it is interpreted as accelerating in a manner consistent with special relativity implying that, at least in this situation, influence can be conceived of as a force.

\end{abstract}

\maketitle

\section{Introduction}
Information-based physics, also known as information physics \cite{Knuth:infophysics}\cite{Goyal:2012information}, is based on the premise that the laws of physics represent the optimal means by which an observer or agent can process relevant information to make predictions about the surrounding world.

In recent works, Knuth and Bahreyni \cite{Knuth+Bahreyni:JMP2014}\cite{Bahreyni:Thesis} have explored to what degree the mathematics of relativistic time and space are derivable from causal interactions.  They demonstrated that partially-ordered sets can be consistently quantified with respect to embedded chains (representing observers), and have proved that in relevant special cases this results in a mathematical formalism that is consistent with special relativistic space-time physics.  That is, the concepts of space and time are emergent in a coarse-grained model of influence events.  Subsequently, Knuth \cite{Knuth:FQXI2013}\cite{Knuth:Info-Based:2014} explored to what degree fermion physics is derivable by considering inferences about such interactions, which is based in part on other recent related foundational studies of probability theory \cite{Knuth&Skilling:2012}\cite{Knuth:ModProbTheory} and quantum mechanics \cite{GKS:PRA}\cite{GK:Symmetry}. It was shown that observer-made inferences regarding the free particle, which is defined as a particle that influences others, but is not itself influenced, result in a situation identical to the Feynman checkerboard model of the Dirac equation \cite{Feynman&Hibbs}.

In this paper, we begin to explore the effect of influence on the emergent properties of a particle.  This initial study suggests that when a particle is influenced, it is interpreted as accelerating in a manner consistent with special relativity implying that, at least in this situation, influence can be conceived of as a force.

\section{Background}

We consider a purposefully simplistic model of interaction based on direct particle-particle influence where pairs of particles\footnote{We use the word `particles' here.  However, we do not really know what a particle is, or whether what this theory refers to as a particle is indeed what we traditionally think of as a particle.  A better word might be `entity' or `entities'.} interact via a directed correspondence.  That is, a single instantiation of an influence-mediated correspondence consists of one particle influencing one other particle.  This allows us to define two \emph{events}: the act of influencing, which is associated with the influencing particle, and the response to being influenced, which is associated with the influenced particle.  It is also assumed that the influence events experienced by a single particle can be ordered\footnote{Such ordering can be introduced by assuming that each particle has a potentially inaccessible internal state that is affected by the influence events.}.
This results in a partially ordered set, or \emph{poset}, where particles are represented by ordered chains of influence events   \cite{Knuth:FQXI2013}\cite{Knuth:Info-Based:2014}, and events are the poset elements.  Some of the details here rely on a basic knowledge of order-theory, which while summarized in \cite{Knuth+Bahreyni:JMP2014}, is covered in more detail in introductory texts \cite{Davey&Priestley}.

An observer is imagined to possess a precise instrument, which can count events along a given particle's chain much like a clock.  The key question examined by Knuth and Bahreyni \cite{Knuth+Bahreyni:JMP2014} was how could one or more observers describe the universe of interacting particles using only such clocks.  They considered a coarse-grained picture of the poset and demonstrated that any consistent observer-based scheme based only on the numbers labeling the sequence of events along the embedded observer chain is unique up to scale \cite{Knuth+Bahreyni:JMP2014}.

Consider an observer chain $\P$ where the events that define the chain are totally ordered and isomorphic to the set of integers under the usual ordering (<).  This allows one to quantify the events along the chain by labeling (numbering) them with integers 1, 2, 3, $\ldots$.  That is, every element $p_x \in \P$ is quantified by a number, or valuation, $v(p_x)$ where for $p_x \leq p_y$ we have that $v(p_x) \leq v(p_y)$.  We will often overload the symbols so that for an element $p_x \in \P$ we will use $p_x$ to represent both the element and its valuation $p_x \equiv v(p_x)$ leaving the reader to rely on context to discern which meaning is intended.

There may be some events in the poset that influence events on the quantifying observer chain $\P$.  Such events are said to \emph{forward project} to the chain, in such a way that there exists a unique mapping, which we shall call $P$ after the name of the chain, taking an event $x$ to an event $Px$ defined by the least event on $\P$ that includes $x$, $Px = \min\{y | x \leq y \; \mbox{and} \; y \in \P\}$.  Similarly, there may exist a set of events that are influenced by elements of the chain $\P$.  In this case we say that such an event $x$ backward projects to the chain, where the back projection $\bar{P}x$ is defined dually as the greatest element of the chain that is included by $x$, $\bar{P}x = \max\{y | x \geq y \; \mbox{and} \; y \in \P\}$.  When an event $x$ both forward and backward projects to the chain $\P$, it can be quantified by a pair of numbers $(Px,\bar{P}x)$. The result is a chain-based coordinate system that covers part of the poset.

The relationship between events along a chain is represented by a \emph{closed interval}, which for two elements $x, z \in P$, is denoted by $[x,z]_{\P}$ and defined as the set of elements between and including the endpoints: $[x,z]_{\P} = \{y \in \P | x \leq y \leq z\}$.  For example, the interval denoted $[4,7]_{\P}$ along a chain is defined by the set of events $\{4, 5, 6, 7\}$.  Since combining intervals (set union) that share a common endpoint is associative, one can show that any non-trivial scalar measure of the interval must be additive \cite{Knuth+Bahreyni:JMP2014}.  This allows one to uniquely define the \emph{length} of the interval $[x,z]_{\P}$ as $d([x,z]_{\P}) = z-x$.

\begin{figure}[h!]
\centering
	[a]
	\begin{minipage}{0.18\textwidth}
	\includegraphics[width=\textwidth]{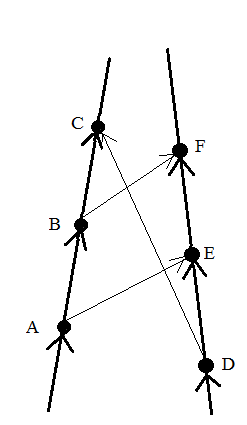}
 	\end{minipage}
 	[b]
	\begin{minipage}{0.27\textwidth}
	\includegraphics[width=\textwidth]{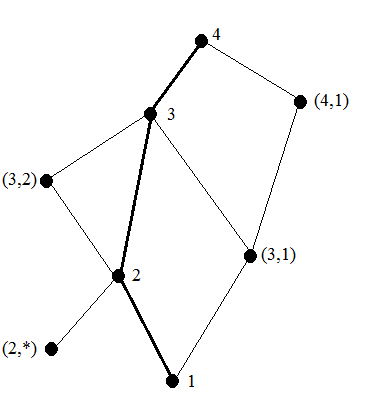}
	\end{minipage}
	[c]
	\begin{minipage}{0.135\textwidth}
	\includegraphics[width=\textwidth]{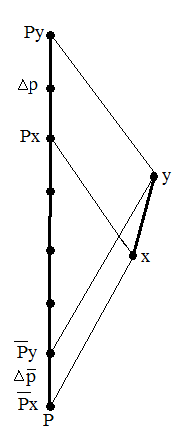}
	\end{minipage}
	[d]
	\begin{minipage}{0.18\textwidth}
	\includegraphics[width=\textwidth]{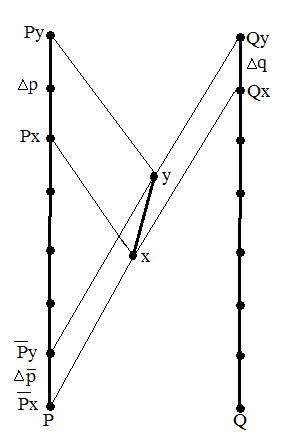}
	\end{minipage}
\caption{\textbf{a:} Events A, B, C are ordered by particle state changes (dark arrows) and so are part of a particle chain. Events D, E, F are on another particle chain. Light arrows are influences. \textbf{b:} The distinguished chain consisting of events labeled $1, 2, 3, 4$ can be used to quantify other events by forward and backward projection resulting in an ordered pair of numbers. \textbf{c:} Interval $[x,y]$ is quantified by chain $\P$ through forward projection of its endpoints to find $ \Delta p $ and back projection to find $ \Delta \bar{p} $. \textbf{d:} Coordinated chains $\P$ and $\Q$ agree on lengths so that $\Delta \bar{p} $ can be replaced by $ \Delta q $.}
\label{fig:Events}
\end{figure}

Observer chains are said to be \emph{compatible} when there exist separate bijectivities between their events under forward and backward projection. Chains $\P$ and $\Q$ are \emph{coordinated} when they are compatible and agree on the lengths of each others' intervals. That is, a closed interval on $\P$ of length $\Delta p$ forward (and backward) projects to a closed interval on $\Q$ with length $\Delta q = \Delta p$.  As a result, the length of such a closed interval can be quantified by the pair of observers by
\begin{eqnarray}
d([p_i, p_j]_{ \mathbf{P} } ) = \frac{\Delta p + \Delta q}{2}
\end{eqnarray}
where $\Delta p = p_j - p_i$, $\Delta q = Qp_j - Qp_i$, and by coordination $\Delta p = \Delta q$.

One can quantify the relationship between coordinated chains $\P$ and $\Q$ by identifying a consistent quantification that depends on the projection of an arbitrary generalized interval $[p_x,q_y]$ where $p_x \in \P$ and $q_y \in \Q$.  It can be shown that the unique measure of a relationship between the two coordinated chains is given by
\begin{equation}
D(\P,\Q) \doteq D([p_x,q_y]) = \frac{\Delta p - \Delta q}{2}
\end{equation}
up to scale where $\Delta p = Pq_y - p_x$ and $\Delta q = q_y - Qp_x$.  For reasons that will be apparent, this measure $D$ is referred to as \emph{distance}.

In general, the relationship between any two elements $x,y$ in the poset, represented by the generalized interval $[x,y]$ defined by the elements, can be quantified by coordinated observer chains $\P$ and $\Q$ if either the forward or backward projections of the elements onto both chains exist. In that case, it can be shown that the interval can be quantified by $\Delta s^2 = \Delta p \Delta q$, which can be decomposed in terms of both length and distance.

By a change of variables we can write length and distance as
\begin{eqnarray}
\Delta t &= \frac{\Delta p + \Delta q}{2} \label{eq:time} \\
\Delta x &= \frac{\Delta p - \Delta q}{2}, \label{eq:space}
\end{eqnarray}
and note that the Minkowski metric results: $ ds^2 = \Delta p \Delta q = \Delta t^{2} - \Delta x^{2} $. The mathematics of flat spacetime has emerged from the poset of events as the result of consistent quantification of intervals by coordinated chains. Despite the fact that there is no space-time per se in the poset picture, one can characterize the behavior of a particle by defining  a \emph{velocity} associated with an interval  to be $ \beta = \frac{\Delta x}{\Delta t} $, which can be written as
\begin{equation} \label{eq:beta}
\beta = \frac{\Delta p - \Delta q}{\Delta p + \Delta q}.
\end{equation}

One can describe a particle behavior as above, in terms of intervals defined by influence events, or equivalently via the Fourier dual, defined by the influence rates. Given a finite particle chain with $N$ being the total number of influences sent, the relevant quantities are the rates\footnote{These definitions differ by a factor of $\frac{1}{2}$ from those in \cite{Knuth:Info-Based:2014} in order that they be properties solely of individual observer chains, since each observer receives half the total number of influences.} $ r_{p} = \frac{N}{2 \Delta p} $ and $ r_{q} = \frac{N}{2 \Delta q} $.  It can then be shown that a quantity analogous to  rest  mass emerges as the geometric average of these rates,
\begin{equation} \label{eq:rest-mass}
M = \frac{N}{ 2  \sqrt{\Delta p \Delta q}};
\end{equation}
a quantity analogous to momentum emerges as the half difference of the rates,
\begin{equation} \label{eq:momentum}
P = \frac{r_{q} - r_{p}}{2};
\end{equation}
and a quantity analogous to energy emerges as the average of the rates,
\begin{equation} \label{eq:energy}
E = \frac{r_{p} + r_{q}}{2}.
\end{equation}
This gives the well-known relation, $ M^{2} = E^{2} - P^{2} $, or equivalently $ E^{2} = M^{2} + P^{2} $.

It is possible to relate quantification of intervals along a particle chain with respect to coordinated observers $\P$ and $\Q$ to quantifications with respect to a linearly-related coordinated pair of chains $\P'$ and $\Q'$. If every interval of length $k$ along $\P$ forward- and back-projects to intervals of lengths $m$ and $n$, respectively, along $\P'$, the following relations can be derived from the coordination condition:
\begin{equation} \label{eq:pprime1}
\Delta p' = \sqrt{\frac{m}{n}} \Delta p
\end{equation}
\begin{equation} \label{eq:qprime1}
\Delta q' = \sqrt{\frac{n}{m}} \Delta q.
\end{equation}
It can be shown that one can write the velocity of the primed observers with respect to $\P$ and $\Q$ as
\begin{equation}
v = \frac{m - n}{m + n},
\end{equation}
and that this can be used to relate $\Delta t'$ and $\Delta x'$ to $\Delta t$ and $\Delta x$, resulting in the \emph{Lorentz transformations} \cite{Knuth+Bahreyni:JMP2014}.

The poset of events can be represented by a variant of a Hasse diagram, in which an event is a vertex; an event including another is higher on the diagram; an influence is a light edge connecting exactly two events; and a particle state is a dark edge. Figure 2a is a diagram for particle $\mathrm{\Pi}$ and observers $\P$ and $\Q$. Each interval between two events on $\mathrm{\Pi}$ forward projects to a zero interval on one observer chain and a nonzero interval on the other, making the velocity $\pm 1$, or the speed of light, for each interval on $\mathrm{\Pi}$ from (\ref{eq:beta}). For example, events a and b both forward project to Q1 but forward project to P1 and P2, respectively. The poset in Figure 2a can be represented in terms of a spacetime diagram (Figure 2b) where the particle is seen to \textit{zitter} at the speed of light.

\begin{figure}
	\centering
	[a]
	\begin{minipage}{0.15\textwidth}
		\includegraphics[width=\textwidth]{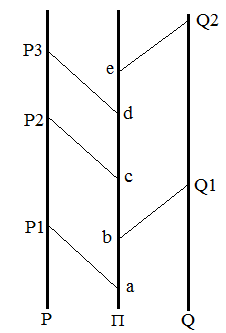}
	\end{minipage}
	[b]
	\begin{minipage}{0.245\textwidth}
		\includegraphics[width=\textwidth]{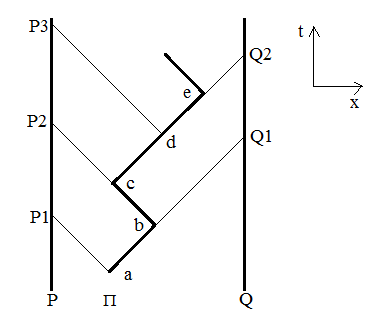}
	\end{minipage}
	\caption{\textbf{a:} A Hasse diagram showing particle $\mathrm{\Pi}$ influencing observer $\P$ three times and observer $\Q$ twice. For example, event c on $\mathrm{\Pi}$ influences $\P$ at event P2. \textbf{b:} This illustrates a path in spacetime consistent with the poset.  Note that the particle zitters back-and-forth at the speed of light.}
	\label{Figure 2}
\end{figure}

If one considers observing a portion of the particle chain for a given amount of time in the spacetime picture, then we find that the particle $\mathrm{\Pi}$ influences $\P$ with probability $Pr(R) = \frac{N_{p}}{N}$, the ratio of the number of P-steps to the total number of steps or the chance of taking a step to the right, and influences $\Q$ with probability $Pr(L) = 1 - Pr(R)$, the probability of taking a step to the left.  These probabilities no longer need to be $\frac{1}{2}$, because observation for a fixed amount of the observers' time is distinct from observation for a fixed amount of the particle's proper time.  Since the influences sent by $\mathrm{\Pi}$ do not change if we introduce different observers $\P'$ and $\Q'$, the probabilities also remain the same.

By (\ref{eq:pprime1}) and (\ref{eq:qprime1}), there exists a frame in which the length of a single P-step and a single Q-step are the same. Since the valuation along chains is determined only up to a scale, we can choose the length of a single step in this frame to be unity, so that $\Delta p = N Pr(R)$ and $\Delta q = N (1 - Pr(R))$ in this frame. In another, primed, frame,
\begin{eqnarray}
\Delta p' &= N Pr(R) \sqrt{\frac{m}{n}} \\
&= \ N Pr(R) k,  \label{eq:pprime2}
\end{eqnarray}
by (\ref{eq:pprime1}), where the symbol $k = \sqrt{\frac{m}{n}}$ has been chosen due to the analogy to Bondi's k-calculus \cite{Bondi:1980} and is the length of one $\P'$ step. Likewise, by (\ref{eq:qprime1}),
\begin{eqnarray}
\Delta q' &= N (1 - Pr(R)) \sqrt{\frac{n}{m}} \\
&= N (1 - Pr(R)) \frac{1}{k}, \label{eq:qprime2}
\end{eqnarray}
where $\frac{1}{k}$ is the length of one $\Q'$ step. By (\ref{eq:rest-mass}), this gives another definition of mass
\begin{equation}
M_{rel} = \frac{1}{ 2  \sqrt{Pr(R)(1-Pr(R))}},
\end{equation}
which can be shown to be analogous to the relativistic mass, which is related to the rest mass $M$ given above in (\ref{eq:rest-mass}) by $M_{rel} = \gamma M$ where $\gamma = (1-v^2)^{-1/2}$.

\section{Fundamental Effects of Influence}

Previous investigations have considered the free particle, which is defined as a particle that influences others, but is not influenced.  Here we present our first results on studying the effect of influence on a particle by considering a particle in 1+1 dimensions that is influenced at a constant rate from a given direction.

\subsection{Acceleration} \label{sec:acceleration}

Consider a particle $ \mathrm{\Pi} $ that not only influences the coordinated observer chains $\P$ and $\Q$ between which it is situated, but also receives influence at a constant rate from $\Q$.
Because $\P$ and $\Q$ are coordinated, they agree on the lengths of intervals projected onto them. The event on the $ \mathrm{\Pi} $ chain representing the receipt of an influence from the
right back-projects to $\Q$.
By coordination of the chains $\P$ and $\Q$, the increment in $ \Delta \bar{q} $ must be matched by an equal increment in $ \Delta p $.
The projected interval length $\Delta \tilde{p}$ in the case where an influence was received is related to the interval length for the free particle $\Delta p$ by
\begin{equation}
\Delta \tilde{p} =  \Delta p + k.
\end{equation}

Since the particle's proper time, $\Delta \tau = \sqrt{\Delta p \Delta q}$, depends only on $ N $, the number of influences that the particle emits to the observers, $ \Delta \tau $ is unchanged by receipt of an influence, giving
\begin{equation}
\Delta \tilde{p} \Delta \tilde{q} = \Delta p \Delta q.
\end{equation}
We can write
\begin{equation} \label{eq:change-to-Delta-p}
\Delta \tilde{p} = \Delta p \frac{\Delta p + k}{\Delta p},
\end{equation}
so that
\begin{equation} \label{eq:change-to-Delta-q}
\Delta \tilde{q} = \Delta q \frac{\Delta p}{\Delta p + k}.
\end{equation}
When the particle emits many more influences to $\P$ for each influence received, we have $\Delta p >> k $, so that we can Taylor expand to find
\begin{equation}
\Delta \tilde{q} \approx \Delta q - \frac{\Delta q}{\Delta p}k.
\end{equation}

The rate $ r $ at which $ \mathrm{\Pi} $ receives influence is defined as
\begin{equation} \label{eq:rconstant}
r \ \dot{=} \ \frac{N _{r} }{N_{p} \Delta \tau},
\end{equation}
where $ N_{r} $ is the constant number of influences received while the particle influences $\P$ $ N _{p} $ times, and $ \Delta \tau$ is the proper time over that interval given by $\sqrt{\Delta p \Delta q}$. This definition of $r$ is motivated by the fact that it will be useful,
since the particle must first influence $\P$, in order to be influenced from the right by $\Q$.\footnote{At present we do not thoroughly understand why this is the case in terms of the poset connectivity, but we have proven this in the case of \textit{Zitterbewegung} using projections of intervals (unpublished).}

Thus for  one  influence received, we have $ \delta \Delta p = k $, and $ \delta \Delta q = -\frac{\Delta q}{\Delta p}k $. The number of these increments in $ \Delta p $ and $ \Delta q $ in proper time $ \Delta \tau $ is $ r N_{p} \Delta \tau $ by
(\ref{eq:rconstant}).
Therefore, the change due to influence is
\begin{equation} \label{eq:dDeltap}
d \Delta p = r k N_{p} \Delta \tau;
\end{equation}
\begin{equation} \label{eq:dDeltaq}
d \Delta q = -r k N_{p} \frac{\Delta q}{\Delta p} \Delta \tau.
\end{equation}
In the expressions above, the proper time $\Delta \tau$ can be considered to be a differential for numbers of events that are large in comparison to unity but small in comparison to those needed to produce times characteristic of the system being considered.\footnote{This is akin to the continuum hypothesis in fluid dynamics.}

From (\ref{eq:pprime2}), we have $ k N_{p} = \Delta p $. The expressions in (\ref{eq:dDeltap}) and (\ref{eq:dDeltaq}) reflect changes due only to received influence. In the absence of received influence, $r \rightarrow 0$, and $\Delta p$ and $\Delta q$ are proportional to
the proper time $\tau$ along the chain,
which yields
\begin{equation}
\frac{d \Delta p}{d \tau} = \left( r + \frac{1}{\tau} \right) \Delta p
\end{equation}
\begin{equation}
\frac{d \Delta q}{d \tau} = \left( -r + \frac{1}{\tau} \right) \Delta q
\end{equation}
as the equations reflecting the effects of both received and emitted influence.

Solutions are
\begin{eqnarray}
\Delta p &= A \tau e^{r \tau}  \\
\Delta q &= B \tau e^{-r \tau}.
\end{eqnarray}
The constants A and B must be reciprocals of each other, since $\Delta p \Delta q = \tau^{2}$. Thus we can write them as $ A = e^{\phi_{0}} $ and $ B = e^{- \phi_{0}} $. We can use the expressions for $\Delta p$ and $\Delta q$ to write an expression for velocity (\ref{eq:beta}) dependent on the proper time
\begin{equation}
\beta = \frac{\exp({r \tau + \phi_{0}}) - \exp({- r \tau - \phi_{0}})}{\exp({r \tau + \phi_{0}}) + \exp({- r \tau - \phi_{0}})},
\end{equation}
so that
\begin{equation}
\beta = \tanh (r \tau + \phi_{0}),
\end{equation}
which is the expression for relativistic velocity under constant acceleration
\cite{Schutz:2009}, with the rate of influence $r$ being identified with the acceleration and $\phi_{0}$ being the initial rapidity.

\subsection{Newton's Second Law} \label{sec:Newton-II}

Now consider the particle receiving influence from the left, at rate $r_{\bar{p}}$, and the right, at rate $r_{\bar{q}}$, where the rates can now be functions of time and position, so long as they are approximately constant over differential increments. By the same arguments as in the previous section, equations (\ref{eq:dDeltap}) and (\ref{eq:dDeltaq}) become

\begin{equation}
d \Delta p = (r_{\bar{q}} - r_{\bar{p}}) d \tau \Delta p
\end{equation}
\begin{equation}
d \Delta q = (r_{\bar{p}} - r_{\bar{q}}) d \tau \Delta q.
\end{equation}

Redefining $r$ as $r \ \dot{=} \ r_{\bar{q}} - r_{\bar{p}}$, we find that (\ref{eq:momentum}) rewritten with a common denominator and the definitions of the rates $r_{p}$ and $r_{q}$ give the change in momentum as
\begin{equation}
dP = \frac{N}{4} \Biggl[ \frac{\Delta p(1 + r d \tau) - \Delta q (1 - r d \tau)}{\Delta p \Delta q} - \frac{\Delta p - \Delta q}{\Delta p \Delta q} \Biggr] .
\end{equation}
Rearranging gives
\begin{equation}
\frac{dP}{d \tau} = \frac{N}{2 \sqrt{\Delta p \Delta q}} \ \frac{\Delta p + \Delta q}{2 \sqrt{\Delta p \Delta q}} \ r.
\end{equation}
The first factor is the rest mass (\ref{eq:rest-mass}), and the second is the ratio of time (\ref{eq:time}) to proper time, which is written in special relativity as $\gamma$. Thus, we can then write
\begin{equation}
\frac{dP}{d \tau} = M \gamma r,
\end{equation}
which is the relativistic version of Newton's Second Law, with the identification of $r$ with acceleration, as found in the previous section\ref{sec:acceleration}.

\subsection{Power}

Under the same conditions as in the previous section\ref{sec:Newton-II}, we can consider the rate of change in the energy of the particle from (\ref{eq:energy}):
\begin{equation}
dE = \frac{N}{4} \Biggl[ \frac{\Delta p(1 + r d \tau) + \Delta q(1 - r d \tau)}{\Delta p \Delta q} - \frac{\Delta p + \Delta q}{\Delta p \Delta q} \Biggr].
\end{equation}
This can be rewritten as
\begin{equation}
\frac{dE}{d \tau} = \frac{N}{2 \sqrt{\Delta p \Delta q}} \ \frac{\Delta p - \Delta q}{\Delta p + \Delta q} \ \frac{\Delta p + \Delta q}{2 \sqrt{\Delta p \Delta q}} \ r.
\end{equation}
The second factor on the right is the velocity (\ref{eq:beta}), and the first and third factors are again the mass and $\gamma$, respectively. By the result of the previous section, $M \gamma r$ is the force, $F$. These identifications enable us to write
\begin{equation}
\frac{dE}{d \tau} = F \beta,
\end{equation}
which is the correct relativistic expression for power.

\section{Conclusion}

In this paper, we have explored the effect of influence on the emergent properties of a particle in 1+1 dimensions.  Our results suggest that when a particle is influenced, it is interpreted as accelerating in a manner consistent with special relativity, which enables one to consider these influences as forces.  We have also shown that this framework allows one to derive the relativistic version of Newton's Second Law as well as the relativistic expression for power.
This is encouraging, since previous work has shown that the Dirac equation can be derived within the same framework by considering a free particle \cite{Knuth:FQXI2013}\cite{Knuth:Info-Based:2014}, which suggests that this picture of emergent spacetime may be consistent with quantum mechanics.  Another paper in this volume \cite{Knuth:MaxEnt2014:motion} by one of us (Knuth) derives the velocity addition law of special relativity and explores the statistical mechanics of motion within this framework. 


\begin{theacknowledgments}
This work was supported, in part, by a grant from the John Templeton Foundation. We wish to thank Newshaw Bahreyni, Ariel Caticha, Seth Chaiken, Philip Goyal, Keith Earle, Oleg Lunin, Anthony Garrett, and John Skilling for interesting discussions and helpful questions and comments.

\end{theacknowledgments}

\bibliographystyle{aipproc}   

\bibliography{knuth}
\end{document}